\documentclass[aps,pra,twocolumn]{revtex4}
\usepackage{amsmath}
\usepackage{amssymb}

\begin{document}
\title{Reply to ``Comment on `Quantum linear Boltzmann equation with finite
       intercollision time' ''}
\author{Lajos Di\'osi}
\email{diosi@rmki.kfki.hu}
\homepage{www.rmki.kfki.hu/~diosi}
\affiliation{
Research Institute for Particle and Nuclear Physics\\
H-1525 Budapest 114, POB 49, Hungary}
\date{\today}

\begin{abstract}
Hornberger and Vacchini [Phys. Rev. A{\bf 82}, 036101 (2010); arxiv:0907.3018] claim 
that the specific collisional momentum decoherence, 
pointed out in my recent work [Phys. Rev. A{\bf 80}, 064104 (2009); arXiv:0905.3908],  
is already described by their theory. 
However, I have performed a calculation whereby I disprove the authors' claim and refute
their conclusion that my recent work had no advantage over theirs.
 
\end{abstract}
\maketitle
In their Comment \cite{HorVac10}, Hornberger and Vacchini (HV) claim without a direct proof 
that the `surprising collisional decoherence effect' (SCDE), central to my new 
quantum linear Boltzmann equation (QLBE) \cite{Dio09}, 
`is fully accounted for in the QLBE' of the authors \cite{VacHor09}. 
If this claim is correct then my work \cite{Dio09} is superfluous.
However, it turns out that the HV QLBE completely ignores
the SCDE. 

Let us study SCDE --- a robust \emph{single collision quantum effect} --- in helium gas 
(molecule mass $m=0.67\times10^{-23}$g, temperature $T=300$K, density $n_g=10^{19}\mathrm{cm}^{-3}$). 
Consider a test particle of mass $M=\mu m$ where $\mu$ is, say, $100$. 
Suppose a cross section $\sigma=10^{-14}\mathrm{cm}^2$ with hard collisions so that 
the momentum transfer be of the order of $\sqrt{mk_BT}$. 
The mean intercollision time becomes $\tau\sim10^{-9}\mathrm{s}$.
Suppose our test particle is in pure state at rest and choose 
the following isotropic coherent momentum spread for it:
\begin{equation}\label{1}
\Delta P=\sqrt{Mk_BT}\sim10^{-17}\mathrm{gcm/s}~.
\end{equation}
Suppose a collision happens to such initial state. According to Eq.(5) of \cite{Dio09}, 
this single collision \emph{completely} decoheres one component $P_\parallel$ of the test 
particle's postcollision momentum. In reality,
there will be a residual coherence $\Delta P_\parallel^{res}$ due 
to the eventual finiteness of both the molecule's coherence length and the intercollision time. 
In our example the latter one dominates. A finite $\tau$ leads
to an uncertainty $\hbar/\tau$ of the energy balance (4) in \cite{Dio09}, 
which allows for a residual postcollision coherence:   
\begin{equation}\label{2}
\Delta P_\parallel^{res}\sim\frac{\hbar}{\tau}\mu\sqrt{\frac{m}{k_BT}}\sim10^{-21}\mathrm{gcm/s}~.
\end{equation}   
This means a reduction of the precollision coherence (\ref{1}) by 4 orders
of magnitude in a single collision. Such sudden drop cannot be resolved by 
kinetic equations, yet a QLBE may qualitatively account for it by an extreme
high value of the coefficient $D_{xx}$ of momentum decoherence. 

The HV QLBE \cite{VacHor09}, as well as Ref.~\cite{Dio95}, predict the momentum decoherence rate 
$D_{xx}(\Delta P)^2$ where 
\begin{equation}\label{3}
D_{xx}=\mathrm{const}\times\frac{1}{\mu\tau Mk_BT}
\end{equation}
is fully classical, $\hbar$ is not involved. 
During a period $\tau$, the predicted ratio of momentum decoherence of the 
chosen initial spread (\ref{1}) takes this simple form:
\begin{equation}\label{4}
D_{xx}(\Delta P)^2\tau=\frac{\mathrm{const}}{\mu}\ll1~.
\end{equation}
The coherence (\ref{1}) looks preserved for times $\sim\mu\tau$ extending over many collisions.
The SCDE \cite{Dio09} is not at all accounted for by
the old QLBE. 
 
My work \cite{Dio09} derives a new expression (23) 
for the coefficient $D_{xx}$, which contains the quantum factor 
$\mathrm{const}\times(\tau k_BT/\hbar)^2\sim10^{8}$ with 
respect to the old $D_{xx}$ (\ref{3}). With $\mu=100$, my QLBE predicts
a decoherence ratio $\sim10^6$ instead of $\sim10^{-2}$ in expression (\ref{4}). This is a heuristic
signal of the SCDE, to indicate that the average drop of the coherence $\Delta P=\sqrt{Mk_BT}$ 
is extremely big on the time scale $\tau$. Once this fast transient
behavior is over, my QLBE is expected to faithfully treat the dynamics of the residual momentum 
coherence in or about the range (\ref{2}). That's the subject of further investigations.  

My QLBE \cite{Dio09} involves new heuristic 
considerations, like the finite time fenomenology of scattering theory. 
Items \emph{(i-iv)} of the HV criticism \cite{HorVac10} may well hold while the criticized
`unfavorable properties' are the price I consciously paid for the SCDE be accounted for.
Item \emph{(v)} is conceptional, but HV's concept of quantum-classical correspondence
should have been made precise, otherwise the related criticism cannot be checked.
Fortunately, the remaining two can be since they criticize physical predictions.
The lack of Gibbs stationary solution \emph{(vi)} is physically 
plausible in my theory where the test particle never ceases to interact with the molecules. 
(I perceive that HV argue against such extension of conservative
scattering theory.)  The corrections with respect to the Gibbs stationary state will all (but Lamb's) 
disappear in the diffusion limit \cite{Dio09} of my QLBE, too. 
Criticism \emph{(vii)} finds unphysical that $D_{xx}$ grows above all bounds when $n_{gas}\rightarrow0$. 
Now, the growth of the SCDE is counterintuitive but real: $\Delta P^{res}$ tends to zero, this needs few
molecules only and a very long $\tau$. Thus real is the growth of $D_{xx}$ as well. Obviously,
my QLBE model of the SCDE will break down before $n_g=0$ and $\tau=\infty$ because, e.g., the residual
coherence $\Delta P^{res}$ might get influenced by the molecules' coherence length.

I have proved that, contrary to the HV Comment, the HV QLBE does not describe 
the SCDE \cite{foot}. I also showed that my QLBE does. HV's criticism has thus been put into a different
perspective. To get SCDE accounted for, I needed more heuristics than before. 

This work was supported by the
Hungarian Scientific Research Fund under Grant No 75129.


\begin{thebibliography}{99}
\bibitem{foot} HV hold a vision of collisional momentum decoherence that somehow differs from mine.
Their vision was likely to prevent them from recognizing the SCDE in its full capacity. Perhaps they 
deny that the residual coherence after a single collision is so little as it in fact is.
Perhaps they agree. It is difficult to determine because the HV Comment does not criticize the SCDE itself.
\end{thebibliography}
\end{document}